\documentclass[prx,aps,twocolumn,superscriptaddress]{revtex4}

\usepackage{dcolumn}% Align table columns on decimal point
\usepackage{bm}% bold math
\usepackage{amssymb}
\usepackage{color}
\usepackage[pdftex]{graphicx}

\usepackage{amsmath}
\usepackage[linesnumbered,ruled]{algorithm2e}

\usepackage{url}
\usepackage[section]{placeins}
\usepackage[colorlinks=true,linkcolor=blue,citecolor=blue]{hyperref}

\begin{document}

%\preprint{Preprint}

\title{Spatio-temporal dynamics of voltage-induced resistance transition \\ in the double-exchange model}

\author{Gia-Wei Chern}
\affiliation{Department of Physics, University of Virginia, Charlottesville, VA 22904, USA}

\date{\today}

\begin{abstract}
We present multi-scale dynamical simulations of voltage-induced insulator-to-metal transition in the double exchange model, a canonical example of itinerant magnet and correlated electron systems.  By combining nonequilibrium Green's function method with large-scale Landau-Lifshitz-Gilbert dynamics, we show that the transition from an antiferromagnetic insulator to the low-resistance state is initiated by the nucleation of a thin ferromagnetic conducting layer at the anode. The metal-insulator interface separating the two phases is then driven toward the opposite electrode by the voltage stress, giving rise to a growing metallic region. We further show that the initial transformation kinetics is well described by the Kolmogorov-Avrami-Ishibashi model with an effective spatial-dimension that depends on the applied voltage. Implications of our findings for the resistive switching in colossal magnetoresistant materials are also discussed.
\end{abstract}

\maketitle

\section{Introduction}

Resistive switching (RS) in a capacitor-like system refers to the drastic changes in resistance induced by a moderate applied voltage or current~\cite{waser07,sawa08,waser09,kim11,jeong12,lee15}. The change of resistance is often non-volatile and reversible. 
The RS effect is not only fascinating by itself, but also has important technological implications, especially for applications in non-volatile information storage, memristor devices, and neuromorphic computing~\cite{zhuang02,baek04,pan14,lee07,valle18}. 
The switching dynamics in real materials is a complex process which involves a large variety of microscopic mechanisms, ranging from thermal effect~\cite{fursina09}, ionic migration~\cite{nian07}, to dielectric breakdown~\cite{lee12} and Mott transition~\cite{cario10}. 
Also importantly, spatial inhomogeneity at the nano-scale plays a crucial role in the resistance transition dynamics. Indeed, extensive experiments have now established that the huge modification of resistance  results from the geometrical transformation of metallic clusters, which could comprise only a small fraction of the driven system in some materials.  Depending on the geometrical pattern of the conducting paths, the RS can be roughly classified into the filamentary-type and the interface-type.

Perhaps the most studied mechanism of RS is the ionic transport facilitated by the electrochemical redox reactions in several oxides. 
In such systems, the switching is controlled by the nano-scale dynamics of ion-migration. % coupled with the redox processes. 
During the so-called electro-forming step, metallic filaments that bridge the two electrodes are formed through electro-chemical reactions. The subsequent ``set" and ``reset" operations correspond to the dissolution  and re-growth, respectively, of the filaments. These mobile ions could be oxygen vacancies already existing in the materials, or cations from the metal electrodes. 
Theoretical modeling of nano-ionic based RS has reached a high level of sophistication. For example, numerical simulations of the ionic filament dynamics ranging from reactive molecular dynamics~\cite{onofrio15,gergs18}, kinetic Monte Carlo~\cite{menzel15,dirkmann15,dirkmann17,guan12}, to continuum diffusion-reaction equation~\cite{kim13,lee13,marchewka16,ambrogio17} coupled with solvers for heat transport and electrostatic potential have been carried out. 
On even larger length scales, effective resistor-network or random circuit breaker models~\cite{chae08,chang09} have been developed to study the statistical and scaling behaviors of filamentary structures~\cite{lee10}.

RS phenomena have also been reported in correlated electron materials in which the switching mechanism is likely of electronic origin. These include the colossal magnetoresistance (CMR) manganites such as La$_{1-x}$Sr$_x$MnO$_3$ (LSMO)~\cite{chen06,krisponeit10,krisponeit13,krisponeit19} and several canonical Mott insulators including vanadium oxides~\cite{vaju08,cario10,janod15,madan15,dubost13}. In particular, RS in the ternary chalcogenides is believed to be driven by Mott insulator-to-metal transition~\cite{dubost13}. Since electron correlation in these materials can be manipulated by various external perturbations such as pressure, temperature, or magnetic field, RS based on correlated electron materials is particularly attractive for multifunctional device applications.

Contrary to the nano-ionic RS, theoretical models of resistance transition in correlated electron systems remain mostly at the phenomenological level. For example, effective resistor network models have been developed to describe the filament structure and dynamics of the switching phenomena~\cite{stoliar14,driscoll12}. While such empirical approaches capture some of the macroscopic features, they do not shed light on the crucial interplay between the microscopic electronic processes and the macroscopic transformation dynamics, hence are limited in their predictive power as quantitative tools for materials design. A comprehensive theory of resistive switching in correlated electron systems thus requires a multi-scale approach that includes the microscopic electronic calculation and the mesoscopic pattern formation simulations.

In this paper, we present the first-ever large-scale dynamical simulation of resistance transition in the double-exchange (DE) model, which is one of the representative correlated electron systems. The DE model describes itinerant electrons interacting with local magnetic moments~\cite{zener51,anderson55,degennes60}. The double exchange mechanism also plays an important role in the CMR phenomena observed in several manganites and diluted magnetic semiconductors~\cite{dagotto01,dagotto_book}.  Since the delocalization of charge carriers requires the alignment of electron spin with the local magnetic moment, the electronic properties of a DE system depends crucially on its magnetic state.  Indeed, the competition between metallic ferromagnetic (FM) clusters and insulating antiferromagnetic (AFM) domains underlies the physics of metal-insulator transition in DE systems. Importantly, the metal-insulator transformation process is controlled with the dynamical evolution of local magnetic moments, with driving force obtained from solutions of the nonequilibrium electron subsystem. 
In order to understand this complex multi-scale phenomenon, we develop a numerical framework that efficiently integrates the nonequilibrium Green's function (NEGF) method~\cite{meir92,jauho94,haug08,datta95,diventra08} with the Landau-Lifshitz-Gilbert (LLG) equation for the spin dynamics. 

\begin{figure}[t]
\includegraphics[width=0.95\columnwidth]{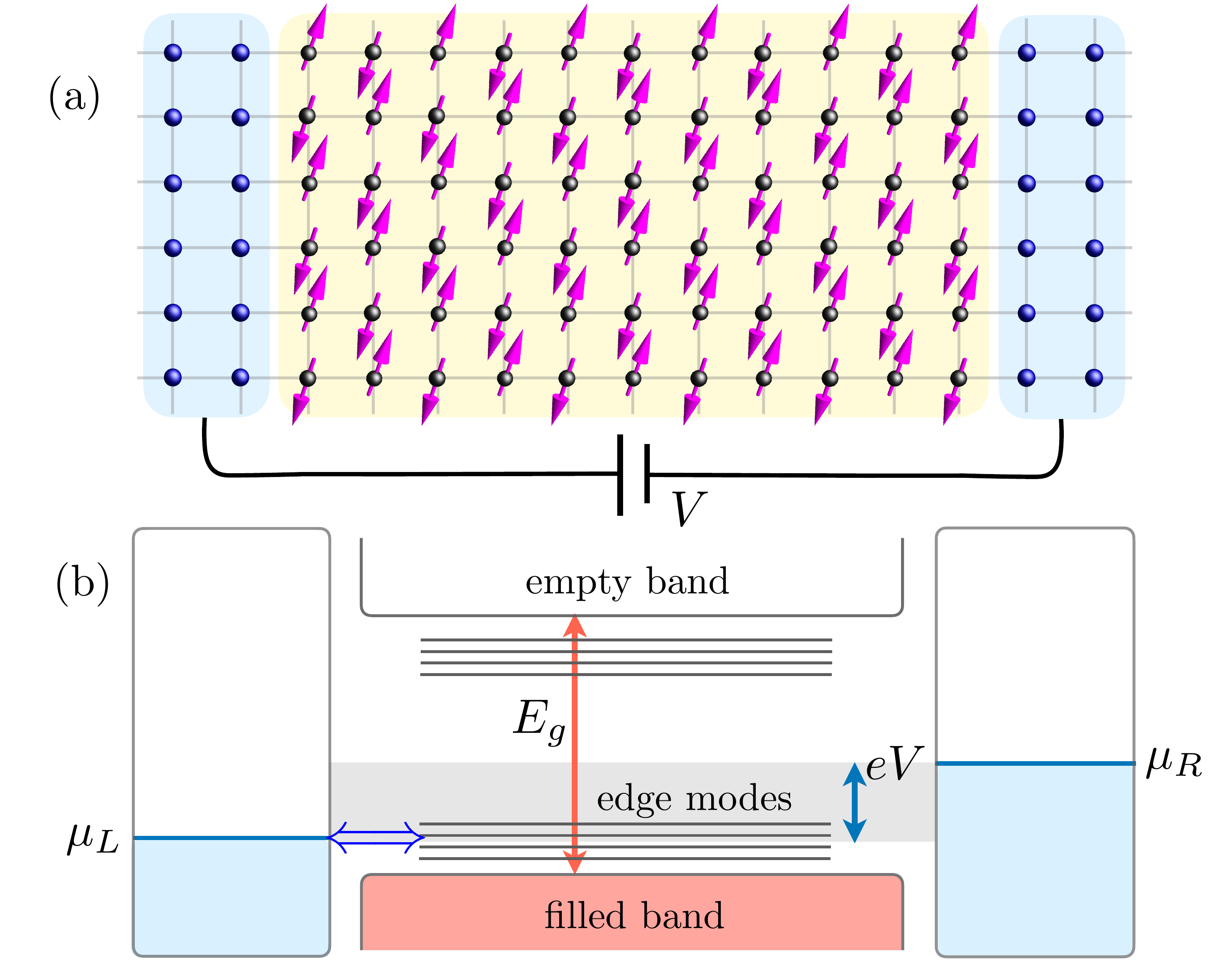}
\caption{(Color online)  
\label{fig:schematic} (a) Schematic diagram of the voltage-driven magnetic transition in the DE model sandwiched by two electrodes. In the initial state, spins in the DE model are arranged in a staggered N\'eel order. A voltage drop $V$ is applied to the two electrodes.  (b) Energy diagram of the system. The DE model in a N\'eel state is a band insulator with an energy gap $E_g = 2 J_{\rm H}$. While the electron chemical potentials of both electrodes lie within the gap, a potential difference is introduced by the voltage: $\mu_R - \mu_L = e V$. 
}
\end{figure}

Here we consider the voltage-induced resistance transition in a DE model sandwiched by two electrodes in a capacitor structure shown in Fig.~\ref{fig:schematic}(a). The DE system is initially in an insulating meta-stable N\'eel state with a large band-gap $E_g$ determined by the electron-spin coupling constant. In the presence of an external voltage that is larger than the bandgap $eV > E_g$, charge carriers at the two electrodes could couple to electron states in the conduction and valance bands of the system. Importantly, due to the delocalized nature of these bulk states, the applied voltage immediately leads to a finite current flow. The bulk of the system quickly becomes unstable and undergoes a fast transformation to a conducting state through a process similar to the dielectric breakdown.

In this work, we instead focus on an insulator-to-metal transition that is induced by a voltage smaller than the band-gap, $eV < E_g$. As shown in Fig.~\ref{fig:schematic}(b), the chemical potentials $\mu_L$ and $\mu_R$ of the two electrodes lie within the band gap in this scenario. Due to the energy mismatch, electrons at the two electrodes cannot efficiently couple to the eigen-states in the bulk. The instability, however, starts at the left edge that is connected to the anode with a chemical potential $\mu_L$ lower than that in the bulk (yet still higher than the valence band edge). Because of the reduced electron density at the edge, anti-parallel spins are unstable against the delocalization of holes through the DE mechanism. The subsequent re-alignment of spins leads to the formation of hole-rich ferromagnetic clusters at the boundary.

The above scenario can also be understood from the energy diagram shown in Fig.~\ref{fig:schematic}(b). The coupling to the two electrodes creates a series of energy-states localized at the two edges. As the applied voltage is increased, the chemical potential $\mu_L$ of the anode is lowered toward the energy levels of the localized modes at the left edge. These edge modes are similar to the band-tail states introduced by disorder. The resultant resonant coupling between electrons at the left electrode and the edge modes leads to an instability toward the formation of a FM layer as electrons are drawn from the sample through the edge modes. The subsequent expansion of the FM domain drives the transformation to the metallic state.  Here we perform the NEGF-LLG simulation to provide a quantitative understanding of the above nucleation and growth scenario of insulator-to-metal phase transformation in the single-band DE model. 

The rest of the paper is organized as follows. In Sec.~\ref{sec:negf-llg}, we discuss the real-space NEGF-LLG method for simulating the adiabatic dynamics of driven DE systems. We next present in Sec.~\ref{sec:imt} the simulation results for the voltage-induced insulator-to-metal transition of a square-lattice DE model.  A detailed analysis of the phase transformation kinetics and the propagation of metal-insulator interface is discussed in Sec.~\ref{sec:kinetics}. Finally, we present a summary and outlook in Sec.~\ref{sec:conclusion}.

 \section{NEGF-LLG dynamics for driven double-exchange systems}

\label{sec:negf-llg}

We consider a two-dimensional capacitor structure, shown in Fig.~\ref{fig:schematic}(a), in which the central region is described by a square-lattice DE Hamiltonian:
\begin{eqnarray}
	\label{eq:H_DE}
	\mathcal{H}_{\rm DE} = - t_{\rm nn} \sum_{\langle ij \rangle}\left( c^\dagger_{i\alpha} c^{\;}_{j\alpha} + \mbox{h.c.} \right)  
	 - J_{\rm H} \sum_i  \mathbf S_i \cdot c^\dagger_{i\alpha} \bm\sigma^{\;}_{\alpha\beta} c^{\;}_{i\beta}, \quad
\end{eqnarray}
where $\langle ij \rangle$ denotes nearest-neighbor pairs, $t_{\rm nn}$ is the nearest-neighbor hopping constant, $J_{\rm H}$ is the on-site Hund's rule coupling constant between local spin $\mathbf S_i$ and conduction electron spin $\mathbf s_i = c^\dagger_{i\alpha} \bm\sigma^{\;}_{\alpha\beta} c^{\;}_{i\beta}$. 
This single-band DE model exhibits several features that are typical of strongly correlated electron systems, such as a filling-controlled insulator to metal transition and electronic phase separation~\cite{yunoki98,dagotto98,chattopadhyay01,luo21}. The DE model could thus  serve as a simple prototype model system to investigate the voltage driven metal-insulator transition in correlated electron materials. Since the DE mechanism is crucial to the CMR effect in manganites~\cite{dagotto01,dagotto_book}, our study here also shed light on RS phenomenon observed in lanthanum manganites such as La$_{1-x}$A$_x$MnO$_3$ (A = Ca, Sr)~\cite{chen06,krisponeit10,krisponeit13,krisponeit19}.

The semiclassical dynamics of local spins in the DE model is governed by the stochastic LLG equation
\begin{eqnarray}
	\label{eq:LL}
	\frac{d \mathbf S_i}{dt}\ =  \mathbf S_i \times \left( \mathbf T_i + \bm\zeta_i \right) - \alpha \mathbf S_i \times \left( \mathbf S_i \times \mathbf T_i \right),
\end{eqnarray}
where $\mathbf T_i$ is the local exchange force, $\alpha$ is the Gilbert damping coefficient, and $\bm\zeta_i(t)$ denotes the stochastic forces described by Gaussian distribution. For equilibrium electronic state, the exchange force is given by the partial derivative of a potential energy: $\mathbf T_i = -\partial E / \partial \mathbf S_i$, where $E = \langle \mathcal{H}_{\rm DE} \rangle = {\rm Tr}(\rho_{\rm eq} \, \mathcal{H}_{\rm DE} )$ is the energy of the quasi-equilibrium electron liquid~\cite{brown63,antropov97,ma11,luo21}. For an out-of-equilibrium quantum state $|\Psi \rangle$, the energy $E$ of the system is not a well-defined quantity. However, the force can still be computed using the generalized Hellmann-Feynman theorem~\cite{diventra00,todorov01,stamenova05,ohe06}, which for the DE model is given by
\begin{eqnarray}
	\label{eq:torque}
	\mathbf T_i = - \langle \Psi | \frac{\partial \mathcal{H}_{\rm DE}}{\partial \mathbf S_i } | \Psi \rangle = J_{\rm H} \, \rho_{i\alpha, i\beta}(\{\mathbf S_i \}) \, \bm\sigma_{\beta\alpha}.
\end{eqnarray}
Here we have introduced the single-particle density matrix $\rho_{i\alpha, j\beta}(t) = \langle \Psi(t) | c^\dagger_{j\beta} c^{\;}_{i\alpha} | \Psi(t) \rangle$. It is worth noting that this electron-induced non-equilibrium force is related to the spin-transfer torque in, e.g. s-d models, and current-induced phenomena such as tunneling magnetoresistance~\cite{salahuddin06,xie17,ellis17,petrovic18,dolui20}.

 The square-lattice DE system is connected to a pair of non-interacting leads at the left and right boundaries. Periodic boundary conditions are assumed in the $y$-direction. Moreover, a bath of non-interacting fermions are coupled to every lattice sites. The total Hamiltonian of our system is $\mathcal{H} = \mathcal{H}_{\rm DE} + \mathcal{H}_{\rm res}$, where $\mathcal{H}_{\rm DE}$ is the DE Hamiltonian in Eq.~(\ref{eq:H_DE}), and the second term describes the electrodes, reservoir degrees of freedom, and their coupling to the DE system:
\begin{eqnarray}
	\label{eq:H_res}
	\mathcal{H}_{\rm res} = \sum_{k, \alpha, i} \varepsilon_k \, d^\dagger_{i, k, \alpha} d^{\;}_{i, k, \alpha} - \sum_{i, k, \alpha} V_{k, i} \bigl(d^\dagger_{i, k, \alpha} c^{\;}_{i, \alpha} + {\rm h.c.} \bigr). \quad
\end{eqnarray} 
Here $d_{i, \alpha, k}$ represents non-interacting fermions from the bath ($i$ inside the bulk) or the leads (for $i$ on the two open boundaries), $\alpha$ is the spin index, and $k$ is a continuous quantum number. For example, $k$ encodes the band-structure of the two leads.

After integrating out the reservoir fermions in both leads and the bath, the retarded Green's function matrix for the central region is $\mathbf G^r(\epsilon) = (\epsilon \mathbf I - \mathbf H - \bm \Sigma^r)^{-1}$, where $\mathbf H$ and $\bm\Sigma^r$ are the matrix representation of the DE Hamiltonian and dissipation-induced self-energy, respectively, in the site-spin space. The explicit matrix elements are
\begin{eqnarray}
	H_{i\alpha, j\beta} =  t_{ij} \delta_{\alpha\beta} - J_{\rm H} \delta_{ij} \mathbf S_i \cdot \bm\sigma_{\alpha\beta},
\end{eqnarray}
\begin{eqnarray}
	\Sigma^r_{i\alpha, j\beta}(\epsilon) = \delta_{ij} \delta_{\alpha\beta} \sum_k \frac{ |V_{i,k} |^2 }{ \epsilon - \epsilon_k + i 0^+ }.
\end{eqnarray}
The resultant level-broadening matrix $\bm\Gamma = i (\bm\Sigma^r - \bm\Sigma^a)$ is diagonal with $\Gamma_{i\alpha, i\alpha} = \pi  \sum_k  |V_{i, k}|^2 \delta(\epsilon - \epsilon_k)$. For simplicity, we assume flat wide-band spectrum for the reservoirs, which leads to a frequency-independent broadening factor with two different values $\Gamma_{\rm lead}$ and $\Gamma_{\rm bath}$.
Next, using the Keldysh formula for quasi-steady state, the lesser Green's function is obtained from the retarded/advanced Green's functions: $\mathbf G^{<}(\epsilon) = \mathbf G^r(\epsilon) \bm\Sigma^{<}(\epsilon) \mathbf G^a(\epsilon)$, and the lesser self-energy is related to the $\Sigma^{r/a}$ through dissipation-fluctuation theorem: 
\begin{eqnarray}
	\Sigma^{<}_{i\alpha, j\beta}(\epsilon) = 2 i \,\delta_{ij}\delta_{\alpha\beta}\, \Gamma_{i} \, f_{\rm FD}(\epsilon - \mu_i). 
\end{eqnarray}
Here $\Gamma_i = \Gamma_{\rm lead}$ or $\Gamma_{\rm bath}$ depending on whether site-$i$ is at the boundaries or in the bulk. The local chemical potential $\mu_i = \mu_0$ for the bath, and $\mu_i = \mu_{L/R} = \mu_0 \mp e V/2$ for the two electrodes, where $V$ is the applied voltage. The transmission current of this nonequilibrium state is   
\begin{eqnarray}
	I = \int d\epsilon \, T(\epsilon) [f_L(\epsilon) - f_R(\epsilon)], 
\end{eqnarray}
where $T(\epsilon) = {\rm Tr}({\bm \Gamma}_R\, {\bf G}^r \, {\bm \Gamma}_L \, {\bf G}^{a} )$ is the transmission function, and $f_{L, R}(\epsilon) = f_{\rm FD}(\epsilon - \mu_{L, R})$ are the Fermi-Dirac distribution functions.

The density matrix $\rho_{i\alpha, j\beta}$, which is required for the force calculation Eq.~(\ref{eq:torque}) in the NEGF-LLG dynamics can now be computed from
\begin{eqnarray}
	\rho^{\;}_{i\alpha, j\beta}\left(\{\mathbf S_i \}\right) =  \int \frac{d\epsilon}{2\pi i} G^{<}_{i\alpha,j\beta}\left(\epsilon; \{\mathbf S_i \} \right),
\end{eqnarray} 
for quasi-steady electron state~\cite{diventra00,todorov01,stamenova05,ohe06}. Here we have explicitly shown the dependence of both the Green's function and the density matrix on the instantaneous spin configuration $\{\mathbf S_i\}$. Given the forces acting on spins, a second-order algorithm is used to integrate the stochastic LLG equation.   In the following, the energy is measured in units of $t_{\rm nn}$, while time is measured in $t_{\rm nn}^{-1}$.

\section{nonequilibrium insulator-to-metal transition}

\label{sec:imt}

\begin{figure*}[t]
\includegraphics[width=1.99\columnwidth]{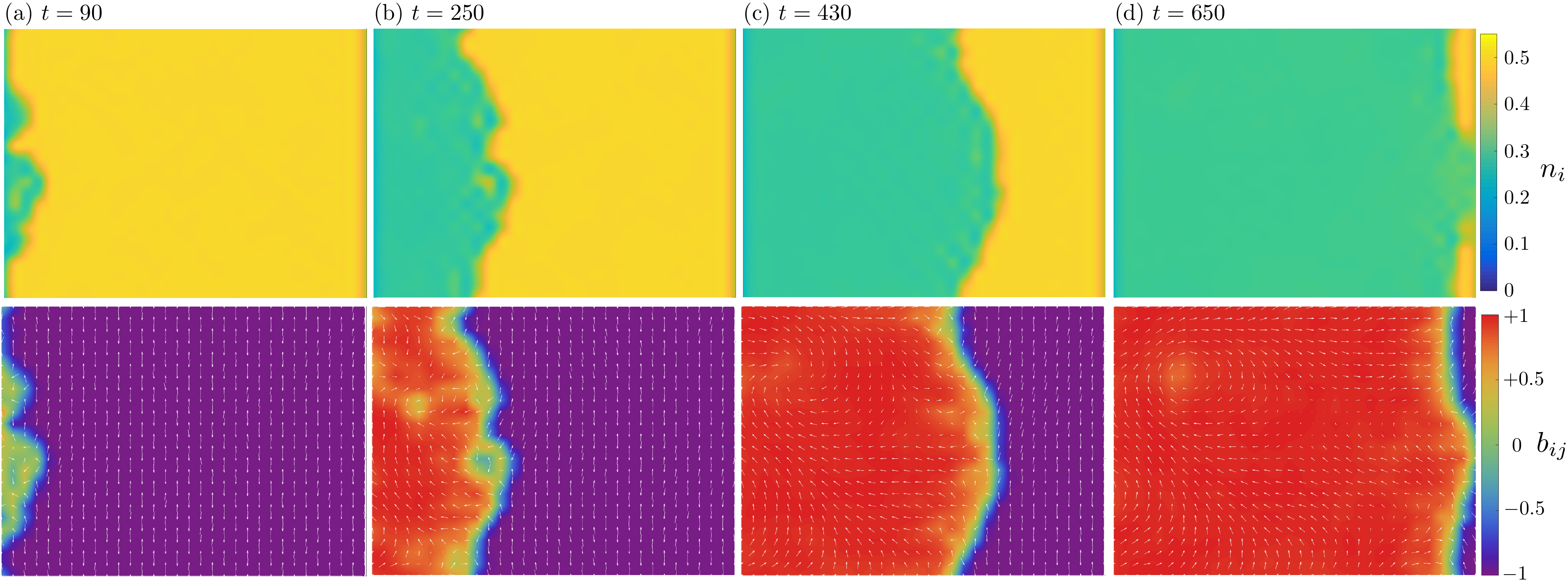}
\caption{(Color online)  
\label{fig:domain-walls} NEGF- LLG simulations of a driven DE model on a $32 \times 24$ square lattice. The voltage bias is applied along the longitudinal $x$ direction. Panels (a)--(d) show the snapshots of the system at various simulation times (in units of inverse nearest-neighbor hopping $t_{nn}$). The top and bottom panels show the local electron density $n_i = \frac{1}{2}\sum_{\alpha} \langle c^\dagger_{i\alpha} c^{\;}_{i\alpha} \rangle$ and the nearest-neighbor spin-spin correlation $b_{ij} = \mathbf S_i \cdot \mathbf S_j$, respectively. The arrows indicate the spin orientations in the $S_x$-$S_z$ plane. $b_{ij} = +1$ ($-1$) corresponds to ferromagnetic (antiferromagnetic) bonds. Simulation parameters: the Hund's coupling $J_{\rm H} = 4.1$, Gilbert damping $\alpha = 0.5$, $\Gamma_{\rm lead} = 1$, $\Gamma_{\rm bath} = 0.01$, $k_B T = 0.0025$, time step $\Delta t = 0.25$, the average $\mu_0 = -3.2$ in the bulk. With a voltage bias $e V = 1.0$, the chemical potential at the left and right leads are $\mu_{R/L} = \mu_0 \pm e V/2$.
}
\end{figure*}

We apply the above NEGF-LLG method to simulate the voltage-induced insulator-to-metal transition of the capacitor structure described by $\mathcal{H} = \mathcal{H}_{\rm DE} + \mathcal{H}_{\rm res}$. The initial state of our simulations was obtained first using the equilibrium LLG dynamics simulations with half-filled electrons $\overline{n} \equiv  \frac{1}{2} \sum_{i, \alpha} n_{i\alpha} / N = 0.5$ per site. As temperature $T \to 0$, this results in a N\'eel order $\mathbf S_i = S (-1)^{x_i + y_i}$ on the square lattice.  The band structure a perfect N\'eel order is given by $E_{\pm}(\mathbf k) = \pm \sqrt{\epsilon_{\mathbf k}^2 + J_{\rm H}^2}$, where $\epsilon_{\mathbf k}$ is the energy dispersion of the square-lattice tight-binding Hamiltonian. Importantly, an energy gap $E_g = 2 J_{\rm H}$ is opened in the spectrum. At half-filling, the valance band $E_-(\mathbf k)$ is completely filled, and the DE system is in a band-insulator state.

Next we turn on the coupling to the bath at a negative chemical potential $0 > \mu_0 > -J_{\rm H}$, which still lies within the gap; see Fig.~\ref{fig:schematic}(b). This coupling can be thought of as a gating-induced doping at high temperatures.  Importantly, this half-filled N\'eel state remains stable as long as $\mu_0$ is within the band gap and the temperature is low enough $k_B T \ll |J_{\rm H} - \mu_0|$.  This stability is confirmed by our NEGF-LLG simulations of the N\'eel state with a reduced chemical potential $\mu_0 = -3.2$ at $V = 0$ and $k_B T = 0.0025$: a thin layer of slightly depleted electrons occurs at each of the two electrodes, while the bulk remains in the half-filled insulating state with almost perfect antiferromagnetic spin order.

In the presence of an external voltage $V > 0$, a chemical potential difference $\Delta \mu = \mu_R - \mu_L = eV$ is introduced between the two electrodes. Crucially, the instability that leads to the insulator-to-metal transition is not driven by this potential difference.  Since the bulk remains gapped at half-filling, the chemical-potential difference $\Delta \mu < E_g$ is not large enough to induce a current flow, which could lead to an instability of the bulk through the DE mechanism.  Instead, the instability comes from the enhanced coupling between the anode and the in-gap modes localized at the left edge when $\mu_L$ significantly overlaps with the energy levels of these edge modes; see Fig.~\ref{fig:schematic}(b).  This resonant coupling between the electrode and the edge modes leads to nucleation of seed ferromagnetic clusters localized at the left boundary. 

%The subsequent transformation to the metallic phase results from the expansion of these ferromagnetic domains with low electron density. 

Fig.~\ref{fig:domain-walls} shows an example of the phase transformation of the driven DE system. The external voltage is turned on at time $t = 0$, giving rise to a chemical potential difference $eV = 1.0$ at the two electrodes. Other simulation parameters are: the Hund's coupling $J_{\rm H} = 4.1$, Gilbert damping coefficient $\alpha = 0.5$, level-broadening coefficients $\Gamma_{\rm lead} = 1$, $\Gamma_{\rm bath} = 0.01$, temperature $k_B T = 0.0025$, and the time step $\Delta t = 0.25$. The chemical potential of the bath is set at $\mu_0 = -3.2$, and the chemical potentials at the two electrodes are $\mu_{R/L} = \mu_0 \pm e V/2$.  The top panels show the snapshots of local electron filling fraction $n_i = \frac{1}{2}\sum_{\alpha} \langle c^\dagger_{i\alpha} c^{\;}_{i\alpha} \rangle$ at different  times of the NEGF-LLG simulations, while the corresponding spin configurations are shown in the bottom panels. 

As discussed above, the nonequilibrium phase transformation starts with the  nucleation of the FM regions at the left edge as electrons are drained to the anode. As the nuclei merge to form a hole-rich domain, a metal-insulator~(MI) interface is created and driven to the right by the voltage stress. Across the MI interface, the electron density $n_i$ changes from $n_e \sim 0.5$ on the insulating side to $n_e \sim 0.3$ on the metallic side. To characterize the spin configurations, we introduce a nearest-neighbor bond variables $b_{ij} = \mathbf S_i \cdot \mathbf S_j$, which is insensitive to the global rotations of spins. The bond-variable serves as an indicator for the short-range spin correlations.  As expected from the DE mechanism~\cite{yunoki98,dagotto98,chattopadhyay01}, FM spin correlation develops in the nucleated hole-rich regions while the insulating domain remains dominated by AFM order; see the bottom panels in Fig.~\ref{fig:domain-walls}.

%Interestingly, this insulator-to-metal transition is not accompanied by a significant increase of the ferromagnetic order parameter. This thus indicates that the ferromagnetic spin-spin correlation of the resultant low-resistance state is short-ranged; e.g. see Fig.~\ref{fig:domain-walls}(d).

\begin{figure}
\includegraphics[width=0.99\columnwidth]{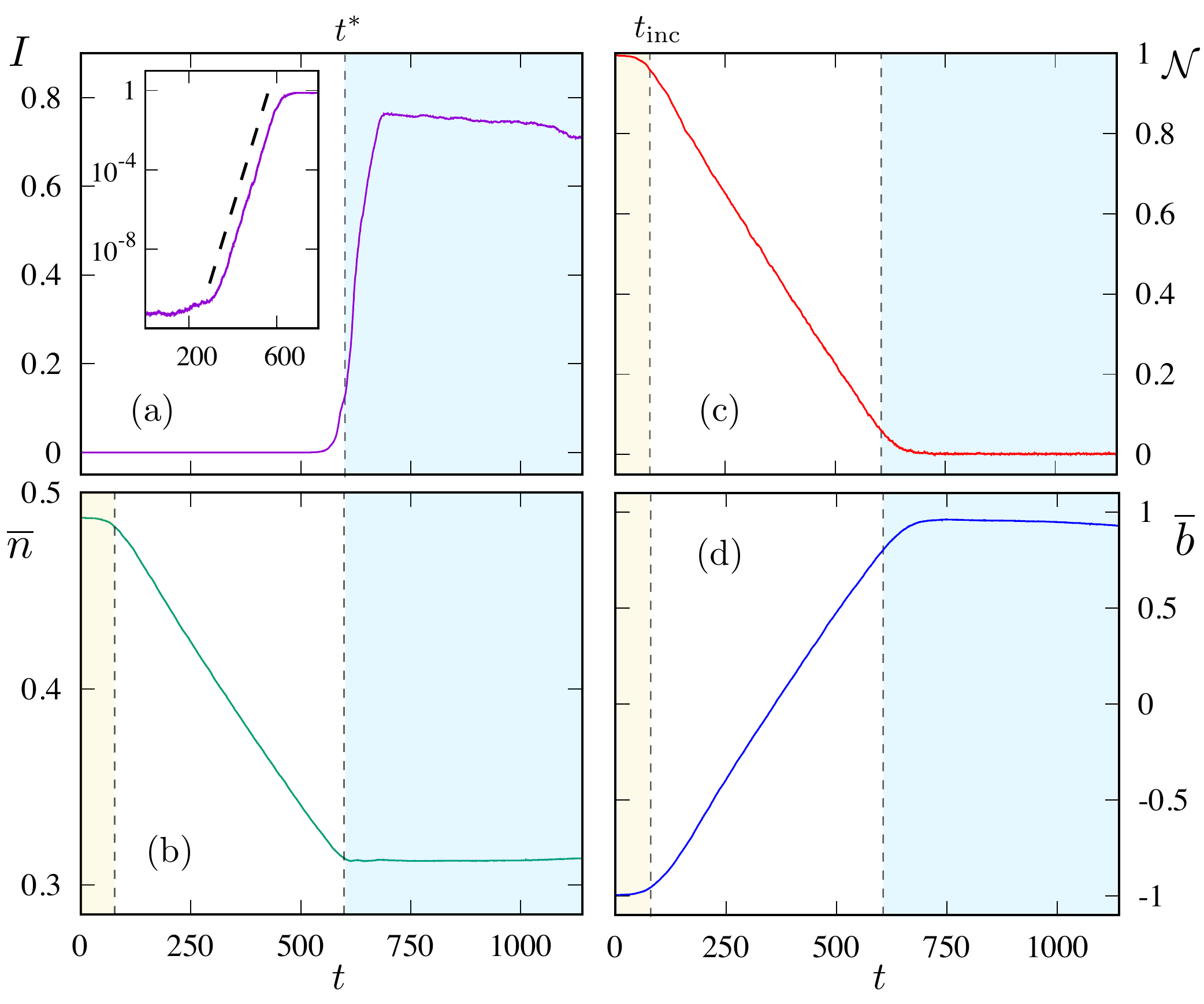}
\caption{(Color online)  
\label{fig:evolution} Time dependence of (a)~average electron filling-fraction $\overline{ n}$, (b) the transmission current $I$,  (c) the antiferromagnetic order parameter $\mathcal{N}$ of spins, and (d)~the average nearest-neighbor spin-correlation $\overline{b} = \frac{1}{2N} \sum_{\langle ij \rangle} b_{ij} $. A constant voltage $eV = 1.0$ is applied to the DE system during the interval $0 \le t \le t_{\rm off}$.   
}
\end{figure}

A more quantitative description of the voltage-driven phase transformation is summarized in Fig.~\ref{fig:evolution} which shows the time dependence of  the transmission current $I$, the spatially averaged electron filling fraction $\overline{n} = \frac{1}{2N}\sum_{i, \alpha} \langle {c}^\dagger_{i\alpha} c^{\,}_{i\alpha} \rangle$, the N\'eel or AFM order parameter $\mathcal{N}\equiv \bigl| \frac{1}{N} \sum_i (-1)^{x_i+y_i} \mathbf S_i \bigr|$, and the spatial-averaged bond variable $\overline{b} =   \frac{1}{2 N}\sum_{\langle ij \rangle} b_{ij}$. 
The initial nucleation of the FM clusters is characterized by an incubation timescale $t_{\rm inc}$. During this period, spins at the left edge gradually realign themselves and form seeds of hole-rich FM  clusters. The AFM order parameter and the electron filling only decrease noticeably for  $t \gtrsim t_{\rm inc}$. 

After the incubation time, a clear metal-insulator (MI) interface is created. During the interval $t_{\rm inc} < t < t^*$, the resultant MI domain-wall propagates across the bulk of the system. As shown in Fig.~\ref{fig:evolution}(a) and (b), the average electron filling fraction $\overline{n}$  decreases almost linearly with time, while the transmission current $I$ remains nearly negligible during this period of domain-wall propagation.  The expansion of the metallic regions also results in a decrease of the AFM order parameter $\mathcal{N}$. The transformation from the AFM to the FM state is also described by the steady increase of the averaged bond-variable $\overline{b}$ from $b_{\rm AMF} = -1$ to $b_{\rm FM} = +1$; see Fig.~\ref{fig:evolution}(c) and (d). As the MI interface reaches the cathode at the opposite end, the FM domain takes over the system, giving rise to a quasi-steady-state regime ($t \gtrsim t^* $) characterized by a nonzero transmission current~$I$ and a low electron filling~$\overline{n} \sim 0.31$.

% We note that when the voltage is turned off, the system remains in the FM metallic state as long as the coupling to the b When the voltage stress $V$ is turned off, the transmission current immediately drops to zero for $t > t_{\rm off}$.  Interestingly, the system remains in a metastable state with a low electron filling fraction and short-range FM spin-spin correlation $\overline{C} \lesssim 1$. Moreover, this metastable state is metallic as coherent propagation of electrons is enabled by the locally parallel spins through the double exchange mechanism.

During the propagation of the MI-interface, the system remains insulating and the transmission current $I$ is very small. Detailed examination shows that the current increases exponentially in this period. The inset of Fig.~\ref{fig:evolution}(a) shows the transmission current $I$ in log-scale as a function of time. The exponential growth also indicates that transmission of electrons at this stage is mainly through quantum tunneling. For an insulating AFM domain of linear size $d$, the tunneling current decays exponentially as the distance: $I \sim e^{-\alpha d}$, where $\alpha$ is a numerical constant. Since the MI-interface propagates with a roughly constant velocity, to be discussed in the following, the thickness of the AFM domain decreases linearly with time $d = L_x - v t$, leading to an exponentially increasing transmission $I \sim e^{+\alpha v t}$.

\begin{figure}
\includegraphics[width=0.99\columnwidth]{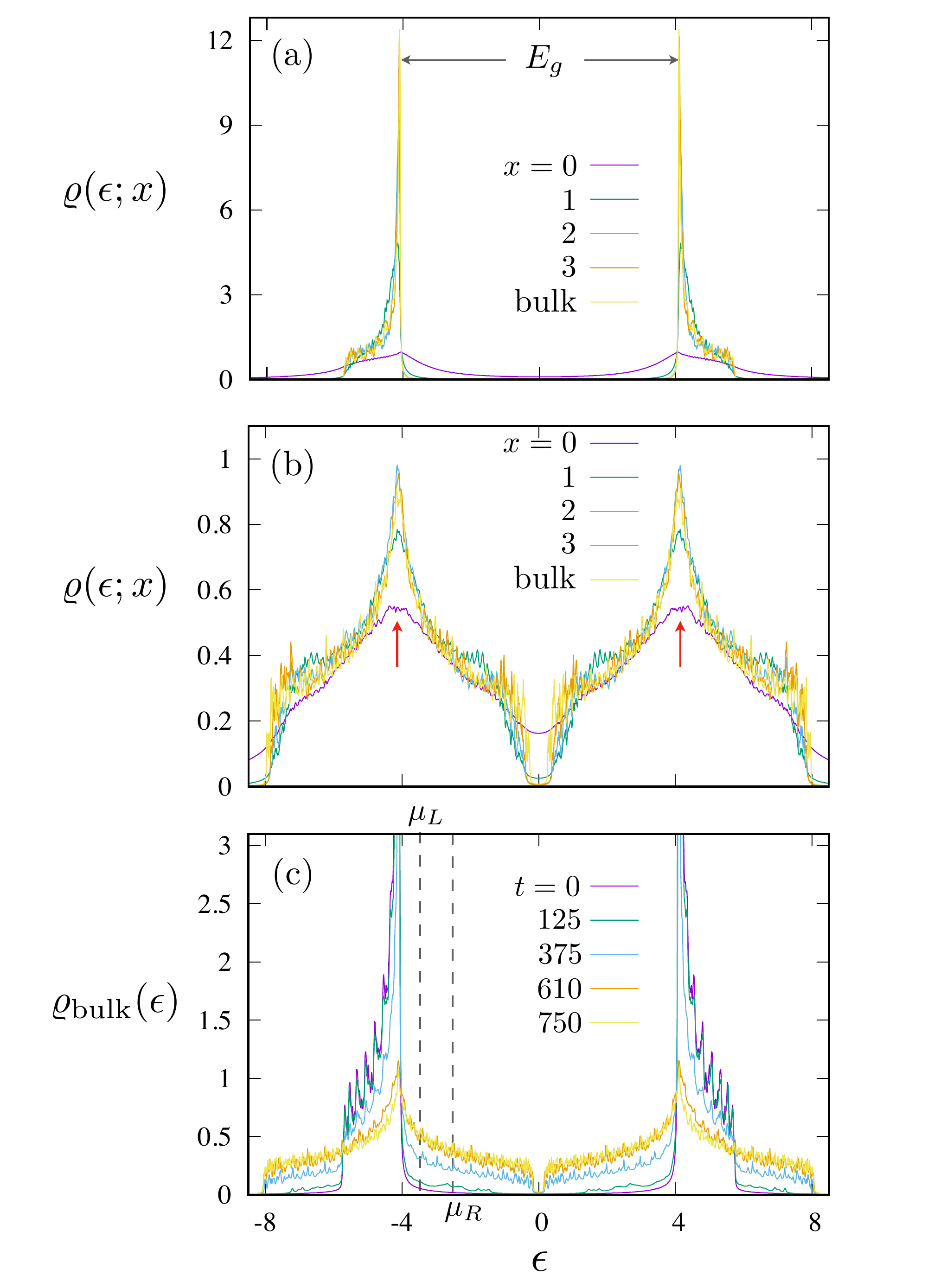}
\caption{(Color online)  
\label{fig:imG} (a) Local density of states (DOS) at $t = 0$, obtained from the imaginary part of the lesser Green's function (averaged over spin and the transverse $y$ direction), for different layers $x$ in the system. The first layer $x = 0$ couples to the left electrode. (b) Spectral function at time $t = 750$ for different layers $x$. In both cases, the yellow curve shows the spectral function averaged over the bulk. The two red arrows correspond to the van Hove singularities of the square-lattice DOS.  (c) The bulk-averaged DOS  at different times during the phase transformation. The two dashed lines indicate the position of the chemical potentials at the left and right electrodes. The Hund's coupling is set at $J_{\rm H} = 4.1$.
}
\end{figure} 

We also compute the local density of states (DOS) from the imaginary part of the electron lesser Green's function
\begin{eqnarray}
	\varrho_i(\epsilon) = \frac{1}{2\pi} \sum_{\alpha} {\rm Im}G^<_{i\alpha,i\alpha}(\epsilon).
\end{eqnarray}
The DOS at various layers near the anode (the left electrode with a lower chemical potential $\mu_L$) at the beginning of the phase transition is shown in Fig.~\ref{fig:imG}(a). Also shown for comparison is the bulk-averaged DOS, which exhibits a pronounced spectral gap $E_g \sim 2 J_{\rm H}$. In fact, the spectral gap is clearly visible even at the second layer.  On the other hand, the leftmost layer ($x = 0$), which couples directly to the electrode, exhibits a broad DOS across the spectrum. This gapless DOS  indicates the ``metallic'' nature of the boundary layer, where the nucleation of the hole-rich FM clusters take places. 

The DOS in the quasi-steady state ($t > t^*$) after the system is transformed into the FM state  is shown in Fig.~\ref{fig:imG}(b). One can understand these spectral functions from the band structure of an FM-ordered DE~system. Here, the dispersion relation $\epsilon_{\mathbf k}$ of the square-lattice tight-binding model is split into two spin-polarized bands: $E_{\pm}(\mathbf k) = \epsilon_{\mathbf k} \pm J_{\rm H}$, where $\pm$ refers to band with electron spins anti-parallel/parallel to the polarized local moments. A small gap $\delta = 2 J_{\rm H} - 8 t_{\rm nn}$ occurs at the origin when the Hund's coupling is greater than half the original bandwidth. The two bands separated by a small gap at $\epsilon=0$ in Fig.~\ref{fig:imG}(b) correspond to the two spin-polarized bands of the FM state, while the two arrows indicate the position of the van Hove singularities of the original square-lattice DOS. 

Fig.~\ref{fig:imG}(c) shows the bulk averaged DOS at different times during the phase transformation. As more and more layers become ferromagnetic, the spectral gap of the AFM state is gradually filled up. The DOS is transformed into two bands with quasi-polarized spins in a state with short-range FM correlation.  The sharp peaks at the band-edges of the AFM state also gradually evolve into two peaks originating from the van Hove singularities of spin polarized bands of the FM state.

\section{Kinetics of phase transformation}

\label{sec:kinetics}

\begin{figure}
\includegraphics[width=0.9\columnwidth]{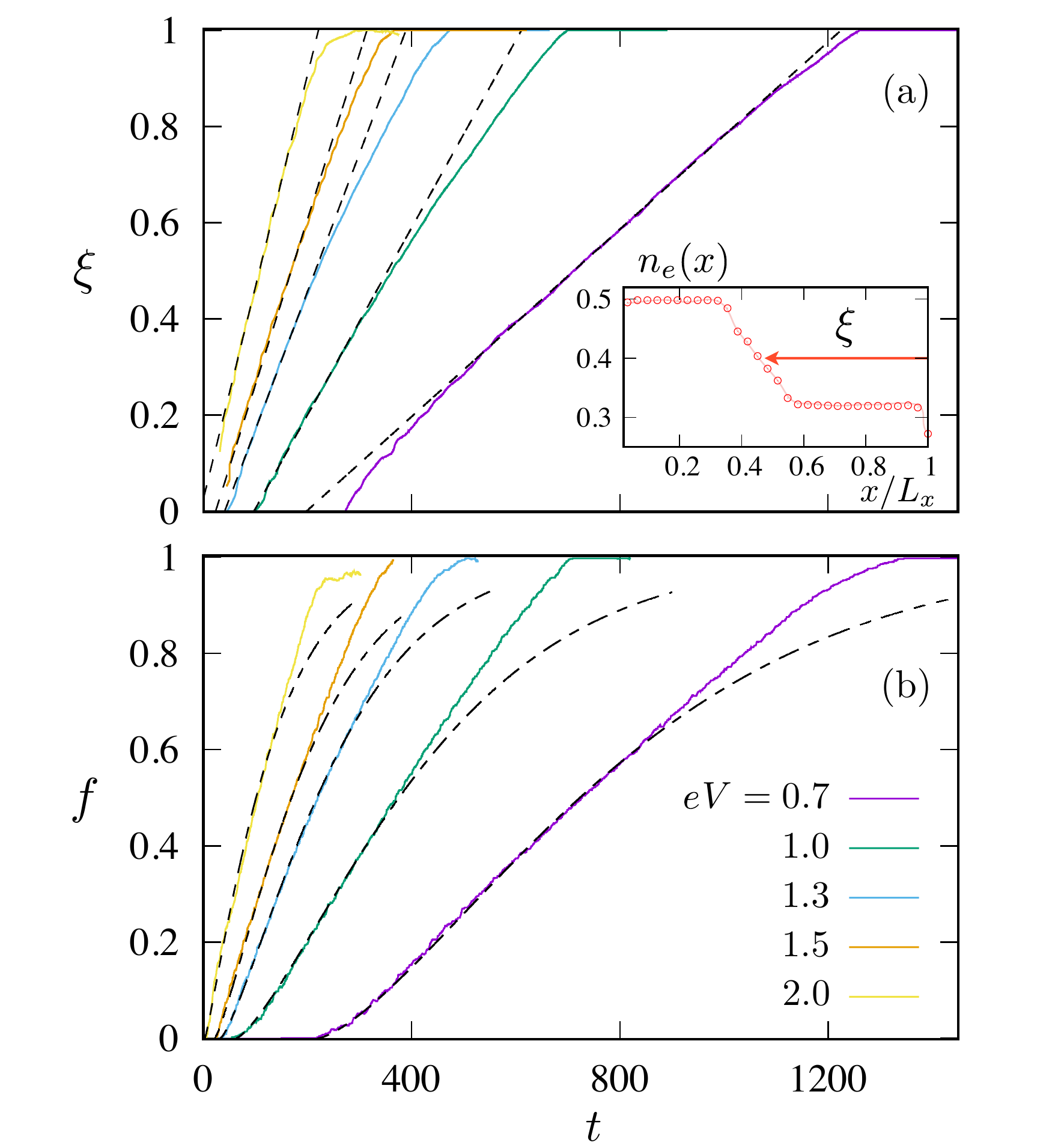}
\caption{(Color online)  
\label{fig:dm-growth} (a) Displacement $\xi$ of MI interface as a function of time for different driving voltages. The dashed lines are fitting to linear function $\xi(t) = v t$ + const. The inset shows the local electron filling $n_e(x)$ (averaged over the transverse $y$-direction) at $t =  750$ for $eV = 0.7$.  (b) Volume fraction $f$ of the transformed FM metallic phase versus time for varying applied voltages. Dashed lines are best fit curves using KAI formulla Eq.~(\ref{eq:KAI}). The DE exchange coupling in both panels is set at $J_{\rm H} = 4.1$.
}
\end{figure}

\begin{figure}
\includegraphics[width=0.95\columnwidth]{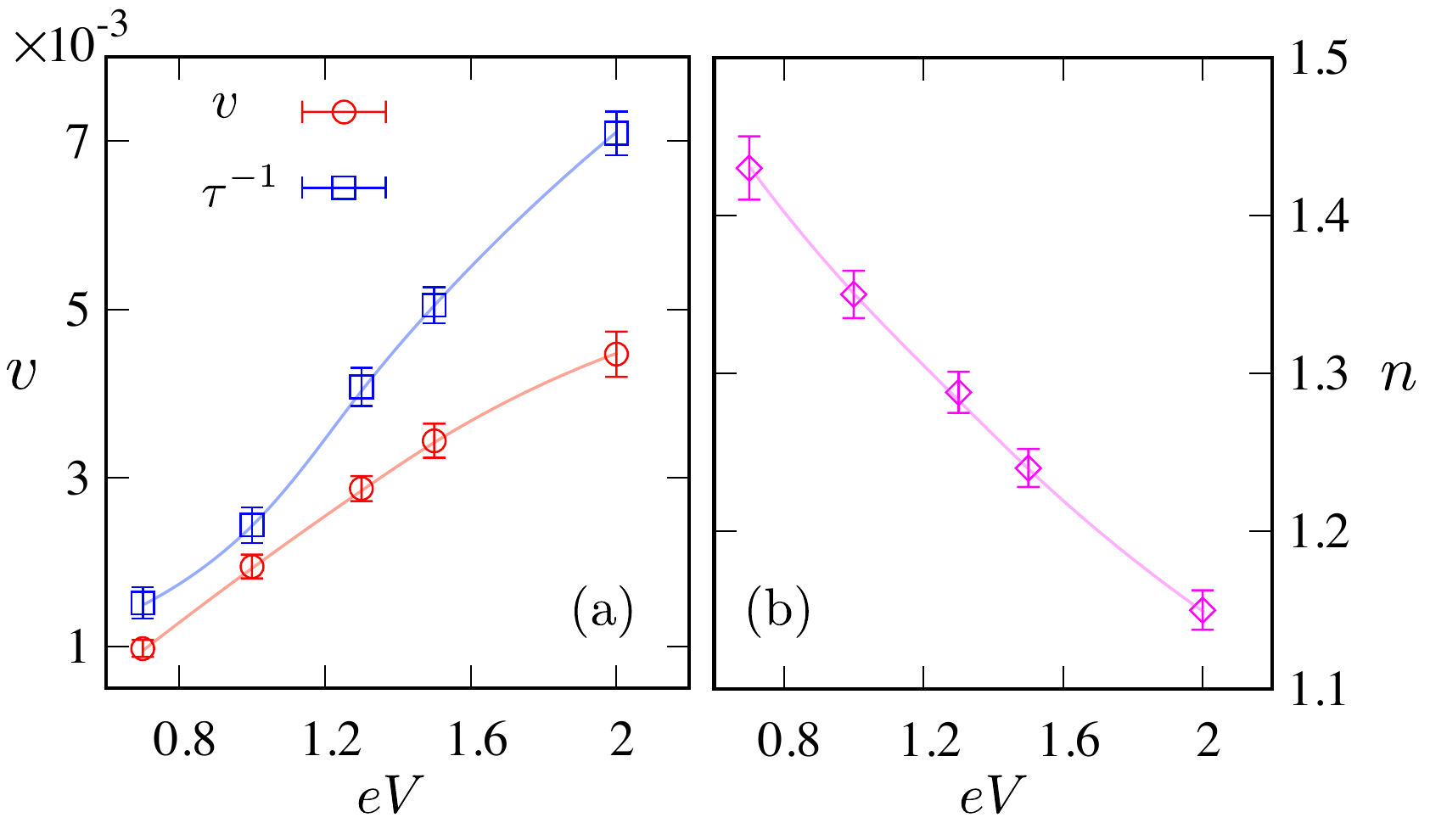}
\caption{(Color online)  
\label{fig:dw-velocity} (a) Velocity $v = d\xi/dt$ of the MI interface as a function of applied chemical-potential difference $\Delta \mu = e V$. Also shown is the inverse characteristic time $\tau^{-1}$ as a function of $eV$. (b) the KAI exponent $n$ as a function of applied voltage.    
}
\end{figure}

Next we turn to the kinetics of the voltage-induced phase transformation. We first consider the propagation dynamics of the metal-insulator (MI) domain walls. To this end, we first compute the electron density profile $n_e(x)$ obtained by averaging $n_i$ over the transverse $y$-direction. An example of the density distribution is shown in the inset of Fig.~\ref{fig:dm-growth}(a). Importantly, the electron density exhibits a sharp discontinuity, which can be used to obtain the position $x_{\rm MI}$ of the MI interface.    Fig.~\ref{fig:dm-growth}(a) shows the (normalized) displacement $\xi = x_{\rm MI} / L_x$ of the MI-interface as a function of time for different driving voltages. Notably, a linear regime characterized by a constant velocity $\xi(t) \sim v t $ emerges after the incubation time $t_{\rm inc}$  that accounts for the nucleation of the FM domains and the formation of the MI interface.   Since quenched disorder is not considered in this work, the motion of the MI-interface is expected to be in the so-called flow regime~\cite{ferrero21,jo09} in which the velocity is proportional to the driving force, $v \sim eV$. This behavior is confirmed in our simulations except for very large applied voltage; see Fig.~\ref{fig:dw-velocity}(a). In the presence of quenched disorder, the propagation of the MI-interface at small voltage is expected to exhibit creep motion and depinning dynamics~\cite{ferrero21}.

To characterize the early stage of the phase transition and particularly the nucleation of the FM domains, we compute the time-dependent volume fraction $f(t)$ of the transformed hole-rich FM phase, which is shown in Fig.~\ref{fig:dm-growth}(b) for varying voltage stresses. These curves are then fitted using the Kolmogorov-Avrami-Ishibashi  (KAI) model~\cite{kolmogorov37,avrami40,ishibashi71}, which has been widely applied to the domain-wall dynamics of ferroelectric transistors~\cite{so05,gruverman05,sharma13,kim17}. In the KAI model, the volume fraction of the transformed phase is described by
\begin{eqnarray}
	\label{eq:KAI}
	f(t) = 1 - \exp\left\{ -\left[ (t - t_{\rm inc}) / \tau \right]^n \right\},
\end{eqnarray} 
where $\tau$ is a time scale characterizing the initial domain growth, $t_{\rm inc}$ denotes the incubation time, and $n$ is called the Avarami exponent. The inverse timescale $\tau^{-1}$ extracted from the fitting is shown in Fig.~\ref{fig:dw-velocity}(a) as a function of voltage. As expected, the time required to grow the proto-FM domain decreases with increasing driving force.

On the other hand, useful information about the growth kinetics is encoded in the Avrami exponent. The original KAI theory predicts that the exponent $n = \alpha + D$~\cite{ishibashi71}, where $D$ is the spatial dimension and the parameter $\alpha = 1$ for  constant nucleation rate, and $\alpha = 0$ for pre-existing nuclei. Since nucleation of the FM-domain only takes place at the electrode, the Avrami exponent is expected to be given by the dimension $D = 2$ in the square-lattice DE model. Interestingly, we find that our NEGF-LLG simulation results can be well described by the KAI mechanism albeit with a non-integer Avrami exponent that decreases with increasing voltage; see Fig.~\ref{fig:dw-velocity}(b). Such fractional effective dimension indicates the nontrivial growth geometry of our system~\cite{ishibashi71,shur98}. One can also understands the voltage dependence as follows. At large voltage stress, the uniform and roughly simultaneous nucleation at the edge of the system results in a unidirectional domain growth and an exponent $n \sim 1$.  On the other hand, the sporadic and non-uniform distribution of the nucleation sites in the case of small voltages give rise to a growth kinetics that preserves some 2D nature of the system as demonstrated in the case of Fig.~\ref{fig:domain-walls}(a).

\section{Conclusion and outlook}

\label{sec:conclusion}

To summarize, we have uncovered surface-induced insulator-to-metal transformation in a double-exchange system driven by an external voltage. The instability of the initial insulating N\'eel state is triggered by the coupling between the electrode and the in-gap modes that are localized at the sample boundary. Our multi-scale NEGF-LLG simulations show that the phase transformation proceeds via the nucleation of a metallic layer at the anode and the subsequent propagation of the metal-insulator domain-wall through the system.  The initial nucleation and growth of the ferromagnetic domains are well described by the KAI model with an effective dimension depending on the voltage stress. The resultant FM-AFM domain wall propagates through the system with an approximately constant velocity, which increases with the driving voltage.  At the end of the phase transformation, the driven DE system enters a nonequilibrium quasi-steady state with a nonzero transmission current. 

The domain-wall driven transition picture is consistent with the experiments on lanthanum manganites La$_{1-x}$A$_x$MnO$_3$ with A = Sr or Ca~\cite{chen06,krisponeit10,krisponeit13}. A non-volatile and bipolar switching with sharp threshold was observed in, e.g. La$_{0.8}$Ca$_{0.2}$MnO$_3$ (LCMO)~\cite{krisponeit10}. Furthermore, using conductive atomic force microscopy, nanometer-sized conducting regions were identified during the RS. Interestingly, the size of the metallic islands was found to depend logarithmically on the pulse-width of the applied electric field, a result that is consistent with the scenario of domain-wall creep and depinning~\cite{lemerle98,chauve00}. This domain-wall propagation picture is further supported by the $1/f^\alpha$ noise of the current during RS~\cite{krisponeit13}, similar to the famous Barkhausen noise in magnetization reversal.  It is worth noting that the above picture is different from the ion-migration controlled interface-type RS suggested for another manganite PCMO~\cite{nian07}. NEGF-LLG simulations for resistance transition in DE systems with quenched disorder will be left for future work. 

Another important generalization is the multi-scale modeling of RS dynamics for Hubbard-type interacting models, which is relevant for RS phenomena in several canonical Mott insulators~\cite{vaju08,cario10,janod15,madan15,dubost13}. It is worth noting that complex spin-density wave patterns have been observed in voltage-driven Hubbard model by solving the self-consistent Hartree-Fock equation with NEGF~\cite{ribeiro16,li17,dutta20}. However, these works only consider static nonequilibrium  solutions of the Hubbard model. A full dynamical modeling of RS phenomena in such systems requires further integration of NEGF-LLG with many-body techniques such as Hartree-Fock method or dynamical mean-field theory. Surrogate many-body solver based on modern machine learning models could be a promising approach for this challenging computational task.

\bigskip

\begin{acknowledgments}
The author thanks A. Ghosh, G. Kotliar, and Jong~Han for fruitful discussions. This work is supported by the US Department of Energy Basic Energy Sciences under Contract No. DE-SC0020330. The authors acknowledge Research Computing at The University of Virginia for providing computational resources and technical support that have contributed to the results reported within this work.
\end{acknowledgments}

\newpage

\end{document}